\documentclass[twocolumn]{aastex631}

\newcommand{\sigmasfr}{$\Sigma_{\rm SFR}$}

\newcommand{\sigmamol}{$\Sigma_{\rm mol}$}

\newcommand{\mout}{$\dot{M}_{\rm out}$}
\newcommand{\tdepout}{$t_{\rm dep,out}$}
\newcommand{\tdepsf}{$t_{\rm dep,SF}$}

\shorttitle{Star Formation Regulation in IRAS08}
\shortauthors{Reichardt Chu et al.}
\graphicspath{{./}{figures/}}

\begin{document}

\title{DUVET: Spatially Resolved Observations of Star Formation Regulation via Galactic Outflows in a Starbursting Disk Galaxy }

\author[0000-0002-7187-8561]{Bronwyn Reichardt Chu}
\affiliation{Centre for Astrophysics and Supercomputing, Swinburne University of Technology, Hawthorn, VIC 3122, Australia}
\affiliation{ARC Centre of Excellence for All Sky Astrophysics in 3 Dimensions (ASTRO 3D)}

\author[0000-0003-0645-5260]{Deanne B. Fisher}
\affiliation{Centre for Astrophysics and Supercomputing, Swinburne University of Technology, Hawthorn, VIC 3122, Australia}
\affiliation{ARC Centre of Excellence for All Sky Astrophysics in 3 Dimensions (ASTRO 3D)}

\author[0000-0002-5480-5686]{Alberto D. Bolatto}
\affiliation{Department of Astronomy, University of Maryland, College Park, MD 20742, USA}

\author[0000-0002-0302-2577]{John Chisholm}
\affiliation{Department of Astronomy, University of Texas, Austin, TX 78712, USA}

\author[0000-0003-3806-8548]{Drummond Fielding}
\affiliation{Center for Computational Astrophysics, Flatiron Institute, 162 Fifth Avenue, New York, NY 10010, USA}

\author[0000-0002-4153-053X]{Danielle Berg}
\affiliation{Department of Astronomy, University of Texas, Austin, TX 78712, USA}

\author[0000-0002-0450-7306]{Alex J. Cameron}
\affiliation{Department of Physics, University of Oxford, Denys Wilkinson Building, Keble Road, Oxford, OX1 4RH, UK}

\author[0000-0002-3254-9044]{Karl Glazebrook}
\affiliation{Centre for Astrophysics and Supercomputing, Swinburne University of Technology, Hawthorn, VIC 3122, Australia}
\affiliation{ARC Centre of Excellence for All Sky Astrophysics in 3 Dimensions (ASTRO 3D)}

\author[0000-0002-2775-0595]{Rodrigo Herrera-Camus}
\affiliation{Departamento de Astronom\'ia, Universidad de Concepci\'on, Barrio Universitario, Concepci\'on, Chile}

\author[0000-0003-1362-9302]{Glenn G. Kacprzak}
\affiliation{Centre for Astrophysics and Supercomputing, Swinburne University of Technology, Hawthorn, VIC 3122, Australia}
\affiliation{ARC Centre of Excellence for All Sky Astrophysics in 3 Dimensions (ASTRO 3D)}

\author[0000-0003-4023-8657]{Laura Lenki\'{c}}
\affiliation{SOFIA Science Center, USRA, NASA Ames Research Center, M.S. N232-12, Moffett Field, CA 94035, USA}

\author{Miao Li}
\affiliation{Department of Physics, Zhejiang University, 866 Yuhangtang Road, Hangzhou, 310058, China}

\author[0000-0001-9345-7234]{Daniel K. McPherson}
\affiliation{Centre for Astrophysics and Supercomputing, Swinburne University of Technology, Hawthorn, VIC 3122, Australia}
\affiliation{ARC Centre of Excellence for All Sky Astrophysics in 3 Dimensions (ASTRO 3D)}

\author[0000-0003-2377-8352]{Nikole M. Nielsen}
\affiliation{Centre for Astrophysics and Supercomputing, Swinburne University of Technology, Hawthorn, VIC 3122, Australia}
\affiliation{ARC Centre of Excellence for All Sky Astrophysics in 3 Dimensions (ASTRO 3D)}

\author[0000-0002-1527-0762]{Danail Obreschkow}
\affiliation{International Centre for Radio Astronomy Research (ICRAR), M468, University of Western Australia,\\ 35 Stirling Hwy, Crawley, WA 6009, Australia}

\author[0000-0001-9719-4080]{Ryan J. Rickards Vaught}
\affiliation{Center for Astrophysics and Space Sciences, Department of Physics, University of California, San Diego, CA, USA}

\author[0000-0002-4378-8534]{Karin Sandstrom}
\affiliation{Center for Astrophysics and Space Sciences, Department of Physics, University of California, San Diego, CA, USA}



\begin{abstract}


We compare 500~pc scale, resolved observations of ionised and molecular gas for the $z\sim0.02$ starbursting disk galaxy IRAS08339+6517, using measurements from KCWI and NOEMA.  We explore the relationship of the star formation driven ionised gas outflows with colocated galaxy properties. 
We find a roughly linear relationship between the outflow mass flux ($\dot{\Sigma}_{\rm out}$) and star formation rate surface density ($\Sigma_{\rm SFR}$), $\dot{\Sigma}_{\rm out}\propto\Sigma_{\rm SFR}^{1.06\pm0.10}$, and a strong correlation between $\dot{\Sigma}_{\rm out}$ and the gas depletion time, such that $\dot{\Sigma}_{\rm out} \propto t_{dep}^{-1.1\pm0.06}$. Moreover, we find these outflows are so-called ``breakout" outflows, according to the relationship between the gas fraction and disk kinematics. Assuming that ionised outflow mass scales with total outflow mass, our observations suggest that the regions of highest $\Sigma_{\rm SFR}$ in IRAS08 are removing more gas via the outflow than through the conversion of gas into stars. Our results are consistent with a picture in which the outflow limits the ability for a region of a disk to maintain short depletion times. Our results underline the need for resolved observations of outflows in more galaxies.

\end{abstract}





\section{Introduction} \label{sec:intro}

Galactic outflows are observed ubiquitously in star-forming galaxies across cosmic time \citep{heckman2000absorptionline, chen2010absorption, steidel2010structure, newman2012sins, Rubin2014evidence, arribas2014ionisedgasoutflows, RodriguezdelPino2019ionisedoutflowsmanga, bolatto2021almaimaging_ngc4945, veilleux2020cooloutflows} and have a critical role in galaxy evolution models \citep[e.g.][]{somerville2015physicalmodels}. When star formation driven outflows are not included, models are unable to reproduce basic galaxy properties such as the galaxy mass function, galaxy sizes, and the Kennicutt-Schmidt Law \citep[see review by][]{somerville2015physicalmodels}. 
The most energetic winds are observed coming from active galactic nuclei \citep{veilleux2005galacticwinds,fluetsch2019coldmolecularoutflows,forsterschreiber2019kmos3d}, which send gas into the halo \citep{nelson2019firstresultsTNG50, oppenheimer2020feedbackSMBH}. 
Galactic fountains arising from clusters of intense star formation remove gas from their local region \citep[e.g.][]{bolatto2013suppression_ngc253,leroy2015almaNGC253,salak2020molecularNGC1482}, which is then recycled into the disk after moving through the lower regions of the halo. 

Star formation driven outflows are widely thought to play multiple roles in regulating star formation. 
One such role is to directly remove gas from regions of active star formation, for example as observed in nearby starbursts NGC~253 \citep{bolatto2013suppression_ngc253} and M~82 \citep{leroy2015multiphaseM82}. In such environments the mass loss rate due to the wind can be comparable to the rate at which gas is converted into stars. 

Moreover, theory argues that the energy and momentum injected from supernovae driven winds plays a regulatory role by increasing the turbulence of the interstellar medium \citep[e.g][]{ostriker2010regulation, fauchergiguere2013feedbackregulatedsf,hayward2017stellar,krumholz2018unifiedmodel}. The increased turbulence is critical to providing pressure support in galaxies, setting the thickness of the gas disk and preventing runaway star formation, and is known as star formation feedback. This regulation generates a local balance between the inward gravitational pressure on the disk, characterised by the molecular and stellar mass surface densities, and the outward pressure provided by the feedback, which is typically characterised by the star formation rate (SFR) surface density multiplied by an efficiency factor \citep[for description see][]{kim2013threeDhydrosim}. Observations do recover positive correlations of gravitational pressure with SFR surface density over 6~orders-of-magnitude \citep{fisher2019testing, barreraballesteros2021Edgecalifa, sun2020dynamicalequilibrium, girard2021systematic}, albeit those published thus far find shallower power-law slopes than theory predicts. \citet{ostriker2022pressureregulated_arXiv} argued that this may be due to systematics in the estimations of pressure. Lenki\'c et al. {\em in prep} will investigate this. It is not clear how outflows may impact feedback regulated star formation theory. Theoretical work employing results from supernova clustering suggest that clustering may lead to increases in the momentum input into the ISM \citep{gentry2020clusteredsupernovae, fielding2018clustered}. Conversely, \cite{orr2022burstingbubbles_letter} argued that if winds leave the disk they carry momentum with them, which might reduce turbulence.   

Current theory therefore implies that supernova driven winds (or outflows) play multiple roles (1) removing gas directly from disks and (2) injecting turbulence into the ISM. Resolved observations of outflows are, however, challenging due to the intrinsic faintness of the spectral features of the wind. 
We use data from the DUVET sample (Fisher et al. {\em in prep}) that uses high signal-to-noise observations from the Keck Cosmic Web Imager (KCWI) to study resolved properties of outflows in rare starburst disk galaxies at $z\approx0.01-0.03$ \citep[e.g][]{cameron2021duvetMrk1486, reichardtchu2022resolvedmaps}. In this paper, we focus on using resolved outflow observations of IRAS~08339+6517 from the DUVET sample to compare the properties of molecular gas and star formation.

\

The paper is organised as follows.  We describe our target galaxy, IRAS~08339+6517 (hereafter IRAS08), in Section~\ref{subsec:IRAS08}. Our observations and data reduction of IRAS08 for ionised gas are described in Section~\ref{subsec:kcwi}, and for molecular gas in Section~\ref{subsec:noema}, together with our methods.  In Section~\ref{sec:stellarmass} we find resolved stellar masses for IRAS08.  In Sections~\ref{subsec:mout}, \ref{subsec:kennicutt_schmidt} and \ref{subsec:outflow_efficiency}  we explore the relationship between outflows and the star formation rate surface density and the molecular gas surface density.  In Sections~\ref{subsec:massloadingfactor} and \ref{subsec:breakout_outflows} we compare our resolved outflow observations to models.  A summary of our results and conclusions is presented in Section~\ref{sec:summary}.
We assume a flat $\Lambda$CDM cosmology with $H_0~=~69.3~\mathrm{km}~\mathrm{Mpc}^{-1}~\mathrm{s}^{-1}$ and $\Omega_0=0.3$ \citep{hinshaw2013wmap9}.

\begin{figure*}
    \centering
    \includegraphics[width=\textwidth]{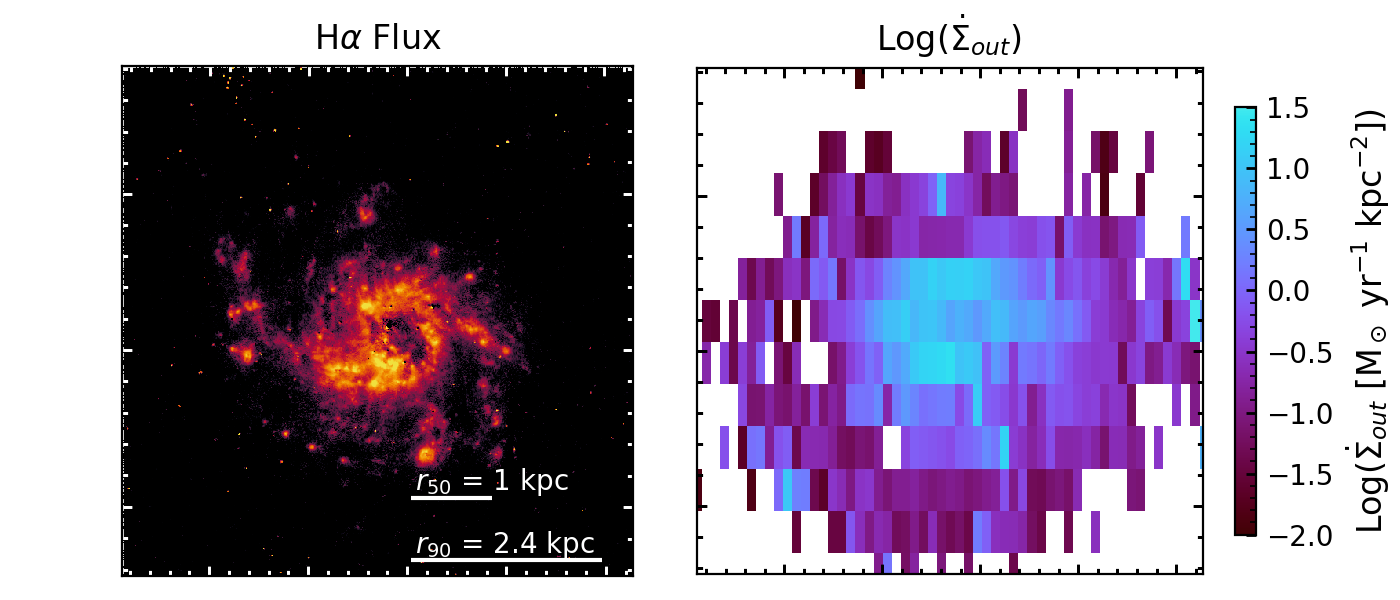}
    \caption{Maps of IRAS08.  (Left) Map of H$\alpha$ flux from {\it HST}. (Right) Mass outflow flux, $\dot{\Sigma}_{\rm out}$, found in each spaxel using [OIII]~$\lambda$5007 and H$\beta$ emission line flux from observations with KCWI on Keck II \citep{reichardtchu2022resolvedmaps}.  Only spaxels where an outflow was detected are included.  The $\dot{\Sigma}_{\rm out}$ values were scaled to assume electron densities of $n_e=100~\mathrm{cm^{-3}}$.  Outflowing gas driven by star formation has been resolved across the disk of IRAS08.}
    \label{fig:maps}
\end{figure*}

\section{Observations and Methods}
\label{sec:observations}

\defcitealias{fisher2022extremevariation}{F22}
\defcitealias{reichardtchu2022resolvedmaps}{RC22}

Our analysis combines observations of outflows from \cite{reichardtchu2022resolvedmaps} (hereafter \citetalias{reichardtchu2022resolvedmaps}) with observations of molecular gas from \cite{fisher2022extremevariation} (hereafter \citetalias{fisher2022extremevariation}). All data have been published, and the data reduction methods are described in more detail in those papers. We briefly summarise these data and methods here.

\subsection{IRAS~08339+6517}
\label{subsec:IRAS08}

IRAS08 is a face-on, UV-bright, blue-compact galaxy at $z\approx0.0191$. It is a $10\times$ outlier from the $z=0$ star forming main sequence.  In Figure~\ref{fig:maps} we show the H$\alpha$ flux map from $HST$, which shows IRAS08's star forming ring.  IRAS08's young stellar population has been shown to be consistent with containing super-star clusters in $HST$/COS spectra \citep{otifloranes2014physicalpropertiesIRAS08}.  For more physical properties of IRAS08, see Table~\ref{tab:IRAS08_properties}.

\begin{table*}[]
    \caption{Galaxy properties for IRAS~08339+6517.}
    \begin{center}
    \begin{tabular}{l c c r}
    \hline
        Property & & Value & Reference \\
        \hline
        Redshift & $z$ & 0.0191 & \citet{kim1995opticalspectroscopy} \\
        Inclination & $i$ & $13^{\circ}$ & \citet{leitherer2013FUVobservations} \\
        Stellar mass & $\mathrm{M}_*$ & $1.1\pm0.3\times10^{10}~\mathrm{M}_\odot$ & \citetalias{fisher2022extremevariation} \\
        Star formation rate & SFR & $12.1\pm1~\mathrm{M}_\odot~\mathrm{yr}^{-1}$ & \citetalias{fisher2022extremevariation} \\
        Stellar population age &  & 10~Myr &  \citet{Leitherer2002globalFUV} \\
        Molecular gas fraction & $f_{\rm gas}$ & 17\% & \citetalias{fisher2022extremevariation} \\
        Toomre stability parameter & $Q_{\rm gas}$ & 0.5 & \citetalias{fisher2022extremevariation} \\
        Resolved ionised gas mass outflow rate ($n_e=100$~cm$^{-3}$) & $\dot{\rm M}_{\rm out}$ & 34.7~$\mathrm{M_\odot~yr^{-1}}$ & \citetalias{reichardtchu2022resolvedmaps} \\
        Resolved ionised gas mass outflow rate ($n_e=300$~cm$^{-3}$) & $\dot{\rm M}_{\rm out}$ & 11.6~$\mathrm{M_\odot~yr^{-1}}$ & \citetalias{reichardtchu2022resolvedmaps} \\
        Total galaxy ionised gas mass loading factor (from absorption) & $\dot{\rm M}_{\rm out}/{\rm SFR}_{\rm COS}$ & 0.07 & \citet{chisholm2017mass}\\
        Resolved ionised gas mass loading factor ($n_e=100$~cm$^{-3}$) & $\eta$ & 3.2 & \citetalias{reichardtchu2022resolvedmaps}\\
        \hline
    \end{tabular}
    \end{center}
    \tablecomments{The mass loading factor $\eta$ from \citetalias{reichardtchu2022resolvedmaps} is calculated using the SFR from only those regions of the galaxy where outflows are observed.}
    \label{tab:IRAS08_properties}
\end{table*}

IRAS08 is well known to host a strong outflow \citep{leitherer2013FUVobservations,chisholm2015scaling,reichardtchu2022resolvedmaps}.  
The benefit of studying IRAS08 is that since it is mostly face-on, we can make point-to-point comparisons between outflow properties and the molecular gas.

\subsection{KCWI observations of SFR and outflows}
\label{subsec:kcwi}

IRAS08 was observed with the Keck Cosmic Web Imager \citep[KCWI,][]{Morrissey2018kcwi} using the BM grating in the Large Slicer mode (FOV: $33\farcs \times 20\farcs4$) for 20 minutes with the central wavelength setting at $\lambda=4800$\AA. These observations are described in both \citetalias{reichardtchu2022resolvedmaps} and \citetalias{fisher2022extremevariation}. The data were reduced using the standard IDL pipeline (Version 1.1.0) and in-frame sky subtraction.   A continuum fit and subtraction was applied using pPXF \citep{cappellari2017ppxf} with BPASS templates \citep[Version 2.2.1][]{stanway2018reevaluatingbpass}.  

To measure SFR and outflow properties we follow a very similar procedure as described in \citetalias{reichardtchu2022resolvedmaps}. The most significant adjustment is that we modify the adopted electron density (discussed below).

KCWI spaxels are non-square, with dimensions $0\farcs29\times1\farcs36$ for the large slicer. \citetalias{reichardtchu2022resolvedmaps} carried out an analysis in which outflows were fit in various spatial bin sizes. They found that correlations of outflow physical properties are more stable for bin sizes of 5$\times$1 spatial sampling of the KCWI data and larger, summing the spectra in each set of 5$\times$1 spaxels.
In order to more easily interpret the data, we resample all data sets to match this 5$\times$1 spatial sampling of the KCWI data. This binning corresponds to sizes of $1\farcs46\times1\farcs36$, which in physical scale is 0.57~kpc$\times$0.53~kpc.

Following the method described in \citetalias{reichardtchu2022resolvedmaps}, to identify outflows we fit both a single and a double Gaussian profile to the H$\beta$ emission lines in spaxels where the continuum has a signal-to-noise greater than 10 per pixel. We used the Bayesian Information Criterion (BIC) to decide whether the extra parameters in the double Gaussian fit are required.  When two Gaussians are required, we assume that the broad second Gaussian represents the outflowing gas (see \citetalias{reichardtchu2022resolvedmaps} for more detail).  

Using the double Gaussian fits, we calculated the outflow velocity
\begin{equation}
    v_{\rm out} = |v_{\rm narrow} - v_{\rm broad}| + 2\sigma_{\rm broad},
    \label{eq:outflow_velocity}
\end{equation}
where $v_{\rm narrow}$ and $v_{\rm broad}$ are the velocities at the centre of the narrow and broad Gaussians respectively, and $\sigma_{\rm broad}$ is the standard deviation of the broad Gaussian.  This definition of $v_{\rm out}$ is comparable to the velocity at 90\% of the continuum, $v_{90}$, measured from absorption line studies \citep[e.g.][]{rupke2005outflowsdiscussion, chisholm2015scaling, chisholm2016shining}, and similar to previous emission line studies \citep[e.g.][]{genzel2011sins, davies2019kiloparsec,fluetsch2019coldmolecularoutflows}. 

The mass outflow rate is defined as
\begin{equation}
    \dot{M}_{\rm out} =  \frac{1.36m_\mathrm{H}}{\gamma_{\mathrm{H}\beta}n_e} \left(\frac{v_{\rm out}}{R_{\rm out}}\right) L_{\rm H\beta,broad},
    \label{eq:mass_outflow_rate}
\end{equation}
where $m_\mathrm{H}$ is the atomic mass of Hydrogen,  
$\gamma_{\mathrm{H}\beta}$ is the H$\beta$ emissivity at electron temperature $T_e=10^4$~K for case B recombination \citep[$\gamma_{\mathrm{H}\beta}=1.24~\times~10^{-25}~\mathrm{erg~cm}^3~\mathrm{s}^{-1}$,][]{osterbrock2006astrophysics}, $n_e$ is the local electron density in the outflow, $R_{\rm out}$ is the radial extent of the outflow,  and $L_{\rm H\beta,broad}$ is the extinction-corrected H$\beta$ luminosity of the broad component. We assume $R_{\rm out}=0.5$~kpc; see the discussion in \citetalias{reichardtchu2022resolvedmaps} for motivation and the likely systematic uncertainty on $R_{\rm out}$.   

\citetalias{reichardtchu2022resolvedmaps} showed that the main driver of $\dot{M}_{\rm out}$ is the luminosity of the broad component, with only a small dependence on the outflow velocity. In IRAS08 there is a $\sim2$ order-of-magnitude variation in the broad-component luminosity, but only a factor of a few change in velocity. 

There is an explicit inverse dependence of $\dot{M}_{\rm out}$ on $\gamma_{\mathrm{H}\beta}$.  Since $\gamma_{\mathrm{H}\beta}$ depends inversely on $T_e$, different assumptions in $T_e$ directly change derived $\dot{M}_{\rm out}$.
\citet{cameron2021duvetMrk1486} provides a direct measurement of T$_e$ from the [OIII]~$\lambda$4363 line for MRK~1486. They found that T$_e$ decreases in the outflow compared to the disk by an order of $\sim$1000~K, with a typical outflow $T_e$ of $\sim$12,000-14,000~K.  
IRAS08 has a higher metallicity than MRK~1486, \citep[0.7~$Z_\odot$,][]{lopezsanchez2006IRAS08paper}, which corresponds to a lower $T_e$  \citep{kewley2019emlinesreview}. Indeed, \cite{lopezsanchez2006IRAS08paper} derives T$_e \sim 8,000$~K for IRAS08. There is no evidence that the metallicity varies across the disk of IRAS08 \citep{lopezsanchez2006IRAS08paper,fisher2022extremevariation}. We, therefore, assume a constant $T_e$ throughout our outflow.
Using the code PyNeb \citep{luridiana2015pyneb} we find that if we assume an electron temperature for the outflow of $T_e=7000$~K, the derived $\dot{M}_{\rm out}$ would be 1.4$\times$ higher than we find for our standard assumption of $T_e=10^4$~K.  Conversely, if we assume a high electron temperature of $T_e=14,000$~K, the $\dot{M}_{\rm out}$ would be 0.7$\times$ lower.

We do not know {\em a priori} what electron density $n_e$ should be adopted for the outflows. 
We only have access to the [OII] density tracer, but the broad outflow components are blended in the spectrum such that we do not have sufficient spectral resolution to measure the electron density of the outflow in IRAS08. We therefore must adopt a value. The electron density of IRAS08 based on the total integrated [OII] emission line ratio from the galaxy and outflow, together, is of order $\sim300-400~\mathrm{cm^{-3}}$. Standard assumptions expect $n_e$ to decline in the outflow compared to the disk, though recent work suggests this may not be the case \citep{forsterschreiber2019kmos3d, fluetsch2021propertiesmultiphase}. 
In the current paper, we are specifically focused on the amount of mass in the outflow. We therefore adopt the electron density scaling as done by \cite{veilleux2002identificationNGC1482}, where the mass outflow is normalised to $n_e=100~\mathrm{cm^{-3}}$. We estimate that $n_e$ introduces a systematic uncertainty of order 0.4-0.5~dex on the outflow mass. In a previous work, \citetalias{reichardtchu2022resolvedmaps} made a different assumption for $n_e$ in order to compare with other observations.  We discuss this further in Sections~\ref{subsec:outflow_efficiency} and \ref{subsec:massloadingfactor}. To illustrate this systematic uncertainty, in Figures~\ref{fig:mout},  \ref{fig:outeff} and \ref{fig:FIRE2_comparison} we include points that represent a higher $n_e$ assumption as well as the $n_e=100~\mathrm{cm^{-3}}$ assumption, as described in associated captions.

To calculate the star formation rate of each spaxel in IRAS08, we use the narrow line flux from the H$\beta$ fits, according to the conversion
\begin{equation}
    {\rm SFR} = \mathrm{C}_{\mathrm{H}\alpha} \left(\frac{L_{\mathrm{H}\alpha}}{L_{\mathrm{H}\beta}}\right) 10^{-0.4A_{\mathrm{H}\beta}} L_{\mathrm{H}\beta}.
    \label{eq:sfr}
\end{equation}
Here C$_{\mathrm{H}\alpha}=5.5335\times10^{-42}$~M$_{\odot}$~yr$^{-1}$~(erg~s$^{-1}$)$^{-1}$, which assumes a \cite{kroupa2003imf} Initial Mass Function \citep{hao2011dustcorrected}.  $L_{\mathrm{H}\alpha}/L_{\mathrm{H}\beta}=2.87$ is the luminosity ratio for electron temperature $T_e=10^4$~K and case B recombination \citep{osterbrock2006astrophysics}. $A_{\mathrm{H}\beta}$ is the extinction derived from the observed H$\beta/$H$\gamma$ ratios and a \cite{calzetti2001dustopacity} attenuation curve.  $L_{\mathrm{H}\beta}$ is the observed H$\beta$ luminosity.

\citetalias{reichardtchu2022resolvedmaps} found the spatially resolved velocity, $v_{\rm out}$, of the ionised gas outflow to be consistent with a shallow slope in $v_{\rm out}\propto\Sigma_{\rm SFR}^{N}$, with $N\propto0.1-0.2$, similar to simulations \citep[e.g.][]{kim2020firstresultssmaug}.  IRAS08 has an integrated mass outflow rate in ionised gas of \mout$\approx7.9~\mathrm{M_\odot~yr^{-1}}$ and a corresponding ionised gas mass loading factor of $\eta_{\rm ion}=\dot{M}_{\rm out}/{\rm SFR}\approx0.8$ when assuming $n_e=380~{\rm cm}^{-3}$ \citepalias{reichardtchu2022resolvedmaps}. In \citetalias{reichardtchu2022resolvedmaps} the integrated mass loading factor $\eta_{\rm ion}$ was calculated using SFRs from only those regions of IRAS08 where evidence of outflows is observed. These values of $v_{\rm out}$, \mout\ and $\eta$ are comparable to strong winds in well studied local galaxies like NGC~253 \citep{bolatto2013suppression_ngc253} and M~82 \citep{shopbell1998asymmetricwindM82}, and suggest that the outflow is removing gas from the galaxy at comparable rates to the star formation.

\subsection{NOEMA CO(2-1) observations of molecular gas}
\label{subsec:noema}

CO(2-1) was observed in IRAS08 using the NOrthern Extended Millimeter Array (NOEMA) for 13 hours in A configuration and 5.5 hours in C configuration.  Observations used the PolyFix correlator tuned to a sky frequency of 226.215~GHz in USB with a channel width of 2.7~km~s$^{-1}$ using 9 antennas.  We recover CO emission over twice the half-light radius of the stars in IRAS08 ($r_{50}\sim1$~kpc or 2.5\arcsec) with a point source sensitivity of 1.4~mJy~beam$^{-1}$ in 20~km~s$^{-1}$ of bandwidth, and with a beam size of $0.52\times0.47$~arcsec$^2$ ($\sim197\times178$~pc$^2$).  For further details on the observations and data reduction, see \citetalias{fisher2022extremevariation}.

To convert CO to H$_2$, we consider both the metallicity and compactness of IRAS08 in selecting the appropriate conversion factor, $\alpha_{\rm CO}$.  IRAS08 has a low metallicity \citep[0.7~$Z_\odot$,][]{lopezsanchez2006IRAS08paper}, indicating that a higher $\alpha_{\rm CO}$ should be used.  On the other hand, IRAS08 is also a compact and starbursting galaxy, suggesting a lower $\alpha_{\rm CO}$. Following the parametrisations from \cite{bolatto2013COtoH2} these two properties offset each other in IRAS08. For simplicity's sake we use a Milky Way $\alpha_{\rm CO}=4.36~\mathrm{M_\odot~(K~km~s^{-1}~pc^2)^{-1}}$ corrected to the CO(2-1) transition using a line ratio of $R_{12}=$CO(2-1)/CO(1-0)=0.7 such that $\alpha_{\rm CO}^{2-1}=\alpha_{\rm CO}/R_{12}=6.23~\mathrm{M_\odot~(K~km~s^{-1}~pc^2)^{-1}}$.
The assumption of $\alpha_{CO}$ introduces a factor of $\sim$2-3 systematic uncertainty in the gas mass surface density that is difficult to characterise further due to the competing effects of the metallicity and the starburst. 

The molecular gas fraction of IRAS08 has been measured to be of order $f_{\rm gas}\equiv M_{\rm mol}/(M_{\rm star} + M_{\rm mol}) \sim 20$\% \citepalias{fisher2022extremevariation}. Toomre's $Q_{\rm gas}$ characterises the stability of a self-gravitating disk, where disks with high velocity dispersion are unstable if $Q_{\rm gas}\leq0.7$ \citep{romeo2010toomrelikestability}.  Dynamically, \citetalias{fisher2022extremevariation} show that IRAS08 is consistent with a galaxy wide violent disk instability \citep{dekel2009formationmassivegalaxies} with $Q_{\rm gas}\sim0.5$ across most of the disk, and high molecular gas velocity dispersions of $\sim25-30$~km~s$^{-1}$.   

\citetalias{fisher2022extremevariation} used NOEMA A+C observations of IRAS08 to study the star formation efficiency per free-fall time, $\epsilon_{\rm ff}\equiv {\rm SFR}/(M_{\rm gas}\times t_{\rm ff})$, where the free-fall time is $t_{\rm ff}\equiv \sqrt{3\pi/(32\mathrm{G}\rho)}$. They found at the 100~pc scale that $\epsilon_{\rm ff}$ reaches high values in the galaxy center of $\sim10-100\%$, which translates to a variable gas depletion time that decreases from $t_{\rm dep,SF}\sim1-2$~Gyr in the disk to $\sim$0.1~Gyr in the galaxy center. The $\epsilon_{\rm ff}$ found for IRAS08 by \citetalias{fisher2022extremevariation} is in the range of values expected from the models of \cite{grudic2018whenfeedbackfails}, which we will discuss further in the context of outflows later in this paper. A strongly variable $\epsilon_{\rm ff}$ is not well explained by current theory, and motivates our comparison to the impact that outflows may have on the star formation regulation.

\subsection{Stellar Mass} 
\label{sec:stellarmass}
We determine the stellar masses in 0.53~kpc$\times$0.57~kpc regions by applying stellar population fits using the CIGALE code \citep{boquien2019cigale} to HST/ACS image filters HRC F330W, WFC F435W, WFC F550M and Spitzer/IRAC Ch1. Before fitting, all stellar continuum images are rotated, convolved to matched PSF and resampled to match the sampling of the KCWI data using Python routines. We remove backgrounds in all images by the standard process of fitting a simple surface to the images in galaxy emission free regions. In the case of the HST images the background is near to zero. We only measure the stellar mass in regions of the image with significant detections of CO(2-1) and the outflow. This restricts the measurement to regions of the stellar continuum images with $S/N\sim50$ or higher in individual resolution elements. 

To carry out the fitting, we assume the \cite{calzetti2000dustcontent} extinction law, allowing for a range of extinctions between $A_V\approx 0.05-2.0$ We note the typical extinction in IRAS08 is quite low, $A_V\approx0.2-0.5$ similar to derivations from the Balmer decrement. We follow the standard prescription for starburst galaxies as outlined in CIGALE papers \citep[e.g.][]{boquien2019cigale}, and fit the star formation history with a delayed starburst superimposed on an exponentially decaying star formation rate. The burst in each pixel is restricted to be less than 15\% of the mass. \cite{ambachew2022stellarmasses} studied the impact of the choice of SFH parametrisation using CIGALE on derived stellar masses in galaxies similar to IRAS08, and find that it has only a small effect. Although we calculate the SFR assuming a Kroupa IMF, CIGALE only implements the Chabrier IMF, which has been shown to under estimate stellar mass-to-light ratios by 5-10\% in the ages that we expect for IRAS08 \citep{conroy2009propogationuncertainties}.  We, therefore, scale the derived masses by 1.08 to match the expected mass-to-light ratio of Kroupa IMFs \citep{conroy2009propogationuncertainties}. We note that this is a minor adjustment compared to the systematic uncertainty in point-to-point measurements of stellar masses in high SFR galaxies. The typical mass-to-light ratio derived from this is of order $M/L_{3.6}\approx0.1-0.3$, which is consistent with similarly young, high SFR surface density  systems \citep{ambachew2022stellarmasses}. We find that the overall fit recovers a stellar mass of $\sim$1.1$\times10^{10}$~M$_{\odot}$ for the entire galaxy, which is similar as reported elsewhere \citep[e.g.][]{leitherer2013FUVobservations,lopezsanchez2006IRAS08paper}. 

\subsection{Prototype Outflow Galaxies For Comparison}

We will compare the molecular gas surface density, $\Sigma_{\rm mol}$, and the star formation rate surface density, $\Sigma_{\rm SFR}$, from IRAS08 to literature values for outflow galaxies in the local Universe (M82, NGC~253 and NGC~1482). For these galaxies throughout this work we adopt values of $\Sigma_{\rm mol}$ and $\Sigma_{\rm SFR}$ that are intended to be from the area of the galaxy near the base of and likely driving the outflow, rather than the global galaxy integrated quantity. For M82, this corresponds to a kiloparsec wide region, which has a molecular gas surface density of $\Sigma_{\rm mol}\sim250$~M$_{\odot}$~pc$^{-2}$ and SFR surface density of $\Sigma_{\rm SFR}\sim$2.5-3~M$_{\odot}$~yr$^{-1}$~kpc$^{-2}$ \citep{leroy2015multiphaseM82,kennicutt1989sflaw}. 

For NGC~253, we use the values given by \cite{leroy2015almaNGC253} for the starbursting nuclear disk, $\Sigma_{\rm mol}\sim560$~M$_{\odot}$~pc$^{-2}$ and $\Sigma_{\rm SFR}\sim2.9$~M$_{\odot}$~yr$^{-1}$~kpc$^{-2}$. 

For NGC~1482, we use the central kiloparsec values from \cite{salak2020molecularNGC1482}, $\Sigma_{\rm mol}\sim490$~M$_{\odot}$~pc$^{-2}$ and $\Sigma_{\rm SFR}\sim1.3$~M$_{\odot}$~yr$^{-1}$~kpc$^{-2}$. 

Note that values of the local outflows (M~82, NGC~253, \& NGC~1482) are only used for context. We never include them in fitting relationships.

\section{Relationship between molecular gas depletion and outflows}
\label{sec:molgasdep_and_outflows}

\subsection{\texorpdfstring{$\dot{\Sigma}_{\rm out}$}{TEXT} and \texorpdfstring{$\Sigma_{\rm SFR}$}{TEXT}}
\label{subsec:mout}

The outflow mass flux is defined as the mass outflow rate normalised by the surface area of the measurement, $\dot{\Sigma}_{\rm out}=\dot{\rm M}_{\rm out}/{\rm Area}$.  This quantity is described in \cite{kim2017tigress}, and is useful for resolved outflow studies. We show the ionised gas outflow mass flux, $\dot{\Sigma}_{\rm out}$, for IRAS08 found using observations with KCWI on Keck II in Figure~\ref{fig:maps}.  Our NOEMA observations were not deep enough to detect broad line emission that could be associated with the molecular component of the outflow.

\begin{figure}
    \centering
    \includegraphics[width=\linewidth]{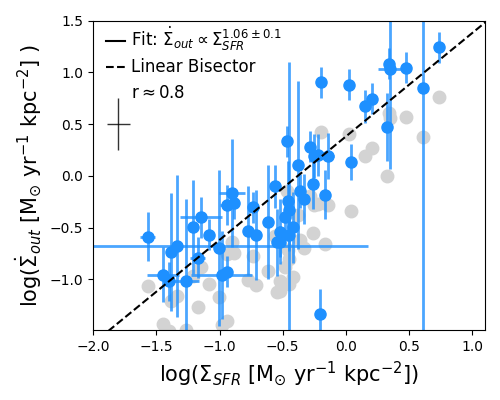}
    \caption{Ionised gas outflow mass flux ($\dot{\Sigma}_{\rm out}$) compared to star formation rate surface density (\sigmasfr) for $\sim500$~pc resolution pixels in IRAS08 where we have observed an outflow. The blue points are calculated assuming an $n_{e}$ normalised to 100~cm$^{-3}$, and the faint grey points use an assumed $n_e$=300~cm$^{-3}$. A typical error bar is shown in the top left corner. We fit a relationship to the two quantities, which returns a slight non-linear correlation. Yet, within the scatter, the data is also consistent with a linear correlation (within 1.5$\sigma$).}
    \label{fig:mout}
\end{figure}

It is commonly expected that a higher star formation rate surface density, \sigmasfr, drives a higher mass outflow rate, \mout\ \citep[e.g.][]{hopkins2012stellarfeedback, hayward2017stellar, arribas2014ionisedgasoutflows, heckman2015systematic, muratov2015gustygaseousFIRE, robertsborsani2020outflowsMANGA}.  In Figure~\ref{fig:mout} we plot $\dot{\Sigma}_{\rm out}$ against \sigmasfr\ for IRAS08 at $\sim500$~pc sampling scale.

In IRAS08 we measure $\dot{\Sigma}_{\rm out}$ $\sim0.1$~M$_{\odot}$~yr$^{-1}$~kpc$^{-2}$  to  $\sim15~\mathrm{M_\odot~yr^{-1}}$~kpc$^{-2}$  over 2 orders of magnitude in \sigmasfr (Fig.~\ref{fig:mout}). We find a Pearson correlation coefficient of $r=0.8$ (p-value $=6\times10^{-13}$) for log(\sigmasfr)-log($\dot{\Sigma}_{\rm out}$). There is a large range in errorbars on $\dot{\Sigma}_{\rm out}$, which would bias the fit to those few points at high signal-to-noise. We, therefore, place a minimum uncertainty of 0.15~dex on $\dot{\Sigma}_{\rm out}$ for the fit.  Using the method of orthogonal distance regression, we find the fit to be
\begin{equation}
    \dot{\Sigma}_{\rm out} = 10^{0.43\pm0.06}~\Sigma_{\rm SFR}^{1.06\pm0.10}.
\end{equation}
In Fig.~\ref{fig:mout} we also show the linear bisector. If we do not constrain the errorbars the fit returns a slightly non-linear powerlaw of $\dot{\Sigma}_{\rm out}\propto \Sigma_{\rm SFR}^{1.12\pm0.09}$.
The root-mean-squared (RMS) deviation of the fit is essentially equivalent to the RMS of the bisecting line, at 0.36~dex. Therefore, while the regression to the unconstrained errorbars returns a slightly nonlinear fit, our results are consistent with a linear relationship $\dot{M}_{\rm out}\propto \Sigma_{\rm SFR}$ (within 1.5$\sigma$).

While there are no observational measurements that we are aware of that compare $\dot{\Sigma}_{\rm out}$ to $\Sigma_{\rm SFR}$, there are some studies that investigate correlations between the integrated star formation and outflow mass rates \citep[e.g.][]{heckman2015systematic}.  We note that there are significant differences between these measurements.  We are measuring $\dot{\Sigma}_{\rm out}$ and $\Sigma_{\rm SFR}$ in ionised gas in equally sized areas in a single galaxy at a fixed distance.  \citet{fluetsch2019coldmolecularoutflows} found a slightly non-linear relationship $\dot{M}_{\rm out}\propto SFR^{1.19\pm0.16}$ for molecular gas outflows in local star forming galaxies. \citet{avery2021incidence} found a similarly linear slope of $\dot{M}_{\rm out}\propto SFR^{0.97\pm0.07}$ for integrated outflows on MaNGA galaxies using H$\alpha$ to measure outflows.

To compare to our results from IRAS08, we calculate the mass outflow rate in ionised gas of M~82 and NGC~1482 
from the total H$\alpha$ luminosity and outflow velocity using values from \cite{shopbell1998asymmetricwindM82} and \cite{veilleux2002identificationNGC1482} 
respectively.  For each of these, we assume, as we do for IRAS08, $n_e\approx100$~cm$^{-3}$ and an $R_{out}=0.5$~kpc.  We find ionised gas mass outflow rates of $\sim$2~M$_\odot$~yr$^{-1}$ for M~82 \citep{shopbell1998asymmetricwindM82} and $\sim$0.6~M$_\odot$~yr$^{-1}$ for NGC~1482 \citep{veilleux2002identificationNGC1482}. 
These \mout\ values translate to $\dot{\Sigma}_{\rm out}\approx 2.5$~M$_\odot$~yr$^{-1}$~kpc$^{-2}$ for M~82,  and $\sim$0.75~M$_\odot$~yr$^{-1}$~kpc$^{-2}$ for NGC~1482. 
We note these are measured in edge-on systems and are thus quite different methods of estimating the mass outflow rate. Nonetheless they are in the range of what we measure for the high \sigmasfr\ regions in IRAS08 (see Fig.~\ref{fig:mout}).

The relationship between $\Sigma_{\rm SFR}$ and $\Sigma_{\rm mol}$ we observe in IRAS08 is broadly consistent with both theoretical expectations and observations of entire galaxies. In the subsequent analysis it will be important to keep track of when the fundamental driver of any relationship may be a reflection of this strong correlation in Fig.~\ref{fig:mout}.

\subsection{Connecting \texorpdfstring{$\Sigma_{\rm mol}-\Sigma_{\rm SFR}$}{TEXT} relationship to outflows}
\label{subsec:kennicutt_schmidt}

\begin{figure}
    \centering
    \includegraphics[width=\linewidth]{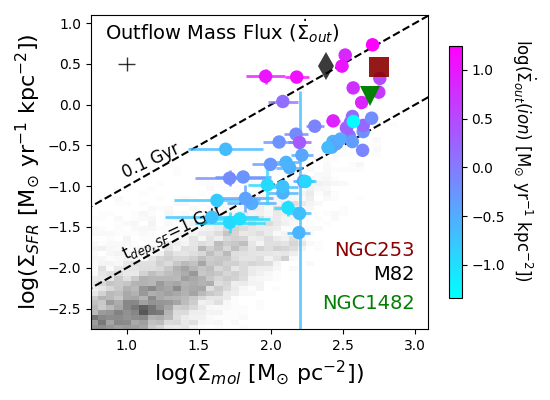}
    \caption{The resolved relationship between star formation rate surface density, \sigmasfr, and molecular gas surface density, \sigmamol. The circles of IRAS08 are coloured according to the mass outflow flux, $\dot{\Sigma}_{\rm out}$. Note that the vertical blue line is the largest of the $\Sigma_{\rm SFR}$ error bars, the rest of the error bars are smaller than the points. A typical error bar is shown in the top left corner.  The grey points represent data from the HERACLES sample of local spirals \citep{Leroy2013molecular}. NGC253 \citep{bolatto2013suppression_ngc253}, M82 \citep{leroy2015multiphaseM82} and NGC1482 \citep{veilleux2002identificationNGC1482,salak2020molecularNGC1482} are represented as a dark red square, a black diamond and a green triangle respectively.  The dashed lines indicate where the time it takes to deplete the molecular gas through star formation, \tdepsf, would be 0.1~Gyr (top line) and 1.0~Gyr (bottom line). In IRAS08, a greater distance from the spiral sequence (grey points) in $\Sigma_{\rm SFR}-\Sigma_{\rm mol}$ corresponds to outflows with higher mass outflow rates and higher momentum flux. }
    \label{fig:kennicutt_schmidt}
\end{figure}

There is a long history of literature discussing the resolved observations of the relationship between \sigmasfr\ and the molecular gas surface density, \sigmamol, in galaxies \citep{kennicutt1989sflaw,bigiel2008sflaw,genzel2011sins,Leroy2013molecular,kennicutt2012star} and theories explaining it \citep{ostriker2010regulation,fauchergiguere2013feedbackregulatedsf,krumholz2018unifiedmodel,hayward2017stellar}.  
\citetalias{fisher2022extremevariation} recently analysed the resolved molecular gas depletion timescale for IRAS08. They find it is consistent with a steep power-law slope for \sigmasfr~$\propto$~\sigmamol$^N$, with $N\approx1.5-1.6$, similar to starburst outflow systems like NGC~253, M~82 and NGC~1482.

In Figure~\ref{fig:kennicutt_schmidt} we show the relation between \sigmasfr\ and the molecular gas surface density, \sigmamol, for IRAS08 in $\sim$0.5~kpc sampling scale.
The high $\Sigma_{\rm SFR}$ center of IRAS08 is comparable in both \sigmasfr\ and \sigmamol\ to other well-known outflow galaxies in the local Universe (NGC~253, M~82 and NGC~1482).

\begin{figure*}
    \centering
    \includegraphics[width=0.8\linewidth]{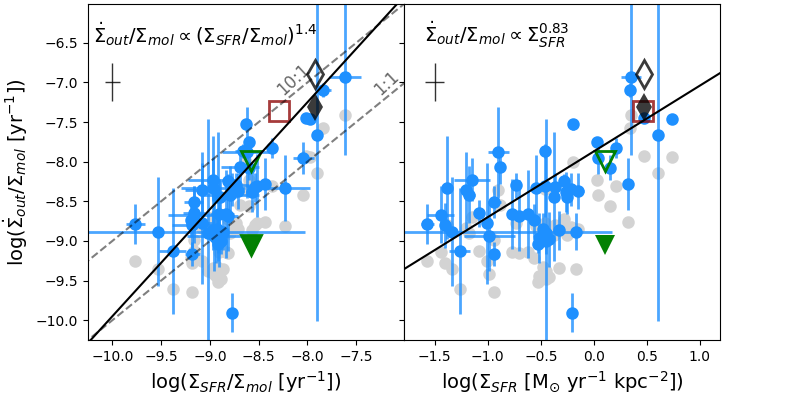}
    \caption{The outflow efficiency $\dot{\Sigma}_{\rm out}/\Sigma_{\rm mol}$ of IRAS08 plotted against the star formation efficiency $\Sigma_{\rm SFR}/\Sigma_{\rm mol}$ (left), and star formation rate surface density \sigmasfr\ (right). In both panels, the blue circles represent our IRAS08 data using $n_e=100~\mathrm{cm^{-3}}$.  Fainter grey circles represent IRAS08 if an electron density of $n_e=300~\mathrm{cm^{-3}}$ were used.  A typical error bar is shown in the top left corner.
    M~82, the central kiloparsec of NGC~1482, and the centre of NGC~253 are the black diamond, green triangle, and dark red square respectively.  Open symbols for these galaxies indicate the results for the molecular gas mass outflow rate, and filled in symbols indicate the ionised gas mass outflow rate.  
    In the left panel, the one-to-one and ten-to-one ratios of the outflow efficiency to star formation efficiency are plotted as grey dashed lines. A fit to the IRAS08 data is given by the solid black line.
    In the right panel, the solid black line shows a fit of the outflow efficiency to $\Sigma_{\rm SFR}$.  
    In high \sigmasfr\ regions, IRAS08 drives more efficient outflows.    }
    \label{fig:outeff}
\end{figure*}

We colour the IRAS08 points by $\dot{\Sigma}_{\rm out}$.
In Fig.~\ref{fig:mout} we showed a tight correlation between $\dot{\Sigma}_{\rm out}$ and $\Sigma_{\rm SFR}$. However, it is clear from Fig.~\ref{fig:kennicutt_schmidt} that there is a large spread in $\Sigma_{\rm mol}$ for a given value of $\dot{\Sigma}_{\rm out}$. The Pearson correlation coefficient is $r=0.8$ (p-value $=6\times10^{-13}$) for log($\dot{\Sigma}_{\rm out}$)-log(\sigmasfr). This reduces significantly to $r=0.5$ (p-value $=4\times10^{-4}$) for log($\dot{\Sigma}_{\rm out}$)-log(\sigmamol).

The average error on the $\Sigma_{\rm SFR}$ points is 0.06~dex.  This typically small value is due to the very high S/N of the KCWI data.  One point in Figures~\ref{fig:mout} and \ref{fig:kennicutt_schmidt} shows a much larger error, and comes from a measurement made with lower S/N than the remainder of the galaxy.

Figure~\ref{fig:kennicutt_schmidt} shows a clear trend that regions of IRAS08 with shorter molecular gas depletion times (indicated by the dashed lines) have larger values of $\dot{\Sigma}_{\rm out}$, 
as indicated by the coloured points. A powerlaw fit to the relationship yields a power-law slope that is close-to-linear, and is likely driven by the correlation in Fig.~\ref{fig:mout}. We find 
\begin{equation}
    \dot{\Sigma}_{\rm out} = 10^{9.2\pm0.5} t_{dep}^{-1.1\pm0.06}
\end{equation}
with a correlation coefficient of $r=-0.85$ and a p-value of roughly $10^{-16}$.

We can also estimate the momentum flux, defined as $dp/dt \equiv v_{\rm out} \dot{M}_{\rm out}$. We do not plot these values, because $dp/dt$ is primarily determined by $\dot{M}_{\rm out}$ and therefore has nearly identical dependency as $\dot{\Sigma}_{\rm out}$. The momentum flux we measure increases from $\sim10^{0.7}~\mathrm{M_\odot~km~s^{-1}~yr^{-1}}$ to $\sim10^{3.2}~\mathrm{M_\odot~km~s^{-1}~yr^{-1}}$. The highest values of $dp/dt$ are similar to what is derived for molecular gas in M~82 \citep{leroy2015multiphaseM82}. 
The correlation for IRAS08 between the momentum flux and the star formation depletion time is significant and non-linear (Pearson correlation coefficient $r=-0.7$, p-value $=2\times10^{-9}$), implying that regions in IRAS08 which are more efficient at turning gas into stars are increasingly more efficient at generating significant momentum in the outflow.

\subsection{Outflow Efficiency}
\label{subsec:outflow_efficiency}

In Figure~\ref{fig:outeff}, we compare $\dot{\Sigma}_{\rm out}/\Sigma_{\rm mol}$, which we refer to as the ``outflow efficiency" of ionised gas, to $\Sigma_{\rm SFR}/\Sigma_{\rm mol}$ and to the unnormalised star formation rate surface density ($\Sigma_{\rm SFR}$). 
The outflow efficiency may be interpreted as the rate at which the outflow exhausts the gas mass with the measured mass outflow rate (assuming there is no supply of fresh gas). We will refer to $\Sigma_{\rm SFR}/\Sigma_{\rm mol}$ as the star formation efficiency for ease of discussion, noting the important distinction between this quantity and the star formation efficiency per free fall time \citepalias[discussed in][]{fisher2022extremevariation}.

In the left panel of Fig.~\ref{fig:outeff} we compare the outflow efficiency to the star formation efficiency. 
When outflow efficiency is greater than star formation efficiency this suggests that outflows more rapidly remove gas from the local region than does the conversion of gas into new stars. We note that the ratio of these two quantities gives the more well-known metric, the mass loading factor $\eta$, which we will consider in Section~\ref{subsec:massloadingfactor}. In IRAS08 the average log($\dot{\Sigma}_{\rm out}/\Sigma_{\rm mol}$ [yr$^{-1}$])$=-8.4$ with standard deviation of 0.55~dex. We find that 87\% of measured regions in IRAS08 fall above the one-to-one line in this space, with a median ratio of $\sim$2.4, when assuming $n_e=100$~cm$^{-3}$. This is similar to unresolved observations of similar quantities, \tdepout\ and \tdepsf, in \cite{fluetsch2019coldmolecularoutflows}. For star formation driven winds in 8 galaxies they found that the galaxies have longer star formation depletion times (smaller SFR/$M_{\rm mol}$) in those galaxies with longer outflow depletion times (smaller $\dot{M}_{\rm out}/M_{\rm mol}$).

We find the outflow efficiency and star formation efficiency are correlated with a Pearson correlation coefficient of $r=0.7$ (p-value $=6\times10^{-10}$). We find the resulting fit, by method of error-weighted orthogonal distance regression, to be 
\begin{equation}
    \frac{\dot{\Sigma}_{\rm out}}{\Sigma_{\rm mol}}= 10^{3.5\pm1.2}\left (\frac{\rm \Sigma_{\rm SFR}}{\Sigma_{\rm mol}}\right )^{1.35\pm0.14}.
\end{equation}
This steeper than linear slope suggests that in regions of high star formation efficiency, the outflow dominates slightly more than it does in regions with star formation similar to that in typical spiral disks. Note that a typical depletion time of $\sim2$~Gyr, as is found in spirals \citep{Leroy2013molecular}, corresponds to a star formation efficiency of $-9.3$~dex in Fig.~\ref{fig:outeff}. It is technically challenging, with present day instrumentation, to observe outflows in typical spirals due to the lower $\dot{M}_{\rm out}$ and smaller velocity offset \citep[e.g.][]{robertsborsani2020outflowsMANGA}. Moreover, in the main disks of spirals 
the ratio of $\Sigma_{\rm SFR}/\Sigma_{\rm mol}$ changes by less than 20\% with galactocentric distance within 8~kpc, not including the galactic nucleus \citep{Leroy2013molecular}. It is, therefore, not clear how this relationship would behave in the disks of spiral galaxies.

We note that the systematic uncertainty introduced from $n_e$ would not necessarily change the trend, however, it may imply that the ionised gas outflow efficiency is more comparable to the star formation efficiency. If the $n_e$ of the outflow changes with $\Sigma_{\rm SFR}$ within a galaxy, this would alter the slope in Fig.~\ref{fig:outeff}, but we do not have any results to suggest whether this occurs. Another large systematic uncertainty, however, is the ratio of the outflow mass flux in different phases, specifically the ionised-to-molecular gas ratio.  
Molecular gas is the more dominant phase of outflow mass \citep[review][]{veilleux2020cooloutflows}, which increases the total \mout\ by a factor of $\sim$5-10$\times$ \citep{fluetsch2019coldmolecularoutflows, herrera-camus2020AGNfeedback}. This suggests that even with a higher $n_e$, outflows likely dominate over star formation in removing the gas in the starbursting center of IRAS08. If the ratio of ions-to-molecules changes with $\Sigma_{\rm SFR}$ this too would affect the relationship between the outflow efficiency and star formation efficiency. In four galaxies, \cite{fluetsch2019coldmolecularoutflows} found a relatively constant relationship between the mass outflow rate measured in ions compared to molecules.  

In the right panel of Fig.~\ref{fig:outeff} we compare outflow efficiency to the SFR surface density. There is likewise a positive correlation between the quantities such that outflows co-located with regions of high $\Sigma_{\rm SFR}$ have higher efficiency. We find a correlation coefficient of $r\approx0.5$ (p-value $=8\times10^{-5}$). We find that, 
\begin{equation}
    \frac{\dot{\Sigma}_{\rm out}}{\Sigma_{\rm mol}}= 10^{-7.86\pm0.05}~\Sigma_{\rm SFR}^{0.83\pm0.11}.
\end{equation}
We can compare this correlation to that of Fig.~\ref{fig:mout}, which has a steeper power law, higher correlation coefficient, and less scatter in the fit as indicated by the fit uncertainties. The larger correlation coefficient implies that the inclusion of the galactic disk molecular gas mass in the outflow efficiency increases scatter in Fig.~\ref{fig:outeff}. 

\begin{figure}
    \centering
    \includegraphics[width=\linewidth]{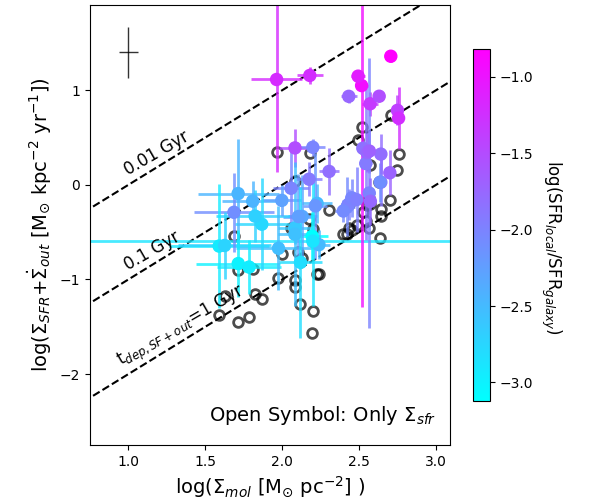}
    \caption{The combined $\Sigma_{\rm SFR}+\dot{\Sigma}_{\rm out}$ is compared to $\Sigma_{\rm mol}$ for regions in IRAS08, and shown as the solid points. Colours indicate the fraction of the total SFR in the corresponding area. A typical error bar is shown in the top left corner. The combined SFR and outflow mass flux represent a more complete view of gas removal. Open symbols represent only $\Sigma_{\rm SFR}$, for comparison. In lower $\Sigma_{\rm SFR}$ regions the combination only makes a factor $\sim$2 change to the gas depletion time (dashed lines), but at high $\Sigma_{\rm SFR}$ it decreases the total depletion time by nearly an order of magnitude. }
    \label{fig:tdep_total}
\end{figure}

In Fig.~\ref{fig:tdep_total} we compare the molecular gas surface density to the combination of both the SFR surface density and outflow mass flux. The combination of $\Sigma_{\rm SFR}+\dot{\Sigma}_{\rm out}$ can be thought of as a more complete estimate of the removal of gas from the star forming region, noting of course that the addition of molecular outflows are needed for a full accounting of the outflow mass. The figure clearly demonstrates that the addition of both the $\Sigma_{\rm SFR}$ and $\dot{\Sigma}_{\rm out}$, together, shortens the depletion time significantly in the high $\Sigma_{\rm SFR}$ regions by nearly an order-of-magnitude. 

Similar to the depletion time measured with only the star formation, the combination of outflow and star formation results in a depletion time that varies within the galaxy and with local $\Sigma_{\rm SFR}$. This is not surprising given our result in Fig.~\ref{fig:mout} that $\dot{\Sigma}_{\rm out} \propto \Sigma_{\rm SFR}$. We find for radii within 1~kpc the median outflow+star formation depletion time ($\Sigma_{\rm mol}/(\dot{\Sigma}_{\rm out} +\Sigma_{\rm SFR})$) is of order 0.05~Gyr. We find a similar value if we select those regions with $\Sigma_{\rm SFR}>1$~M$_{\odot}$~yr$^{-1}$~kpc$^{-2}$. This depletion time increases to  0.3~Gyr for larger radii and lower $\Sigma_{\rm SFR}$. We note that while 1~kpc is a small area, it represents over 50\% of the total star formation in the galaxy, as indicated by the colours in Fig.~\ref{fig:tdep_total}. 

We reiterate that the main systematic uncertainty in our results is that the observations of outflows are only of a single phase, and that the addition of the molecular phase could increase the outflow mass rate by as much as an order-of-magnitude. This would reduce the total depletion time to only 0.005-0.01~Gyr in the center of IRAS08, which is very short, and is comparable to the free-fall time found for these regions \citepalias{fisher2022extremevariation}.

\subsection{Mass-loading Factors}
\label{subsec:massloadingfactor}

\begin{figure*}
    \centering
    \includegraphics[width=0.8\linewidth]{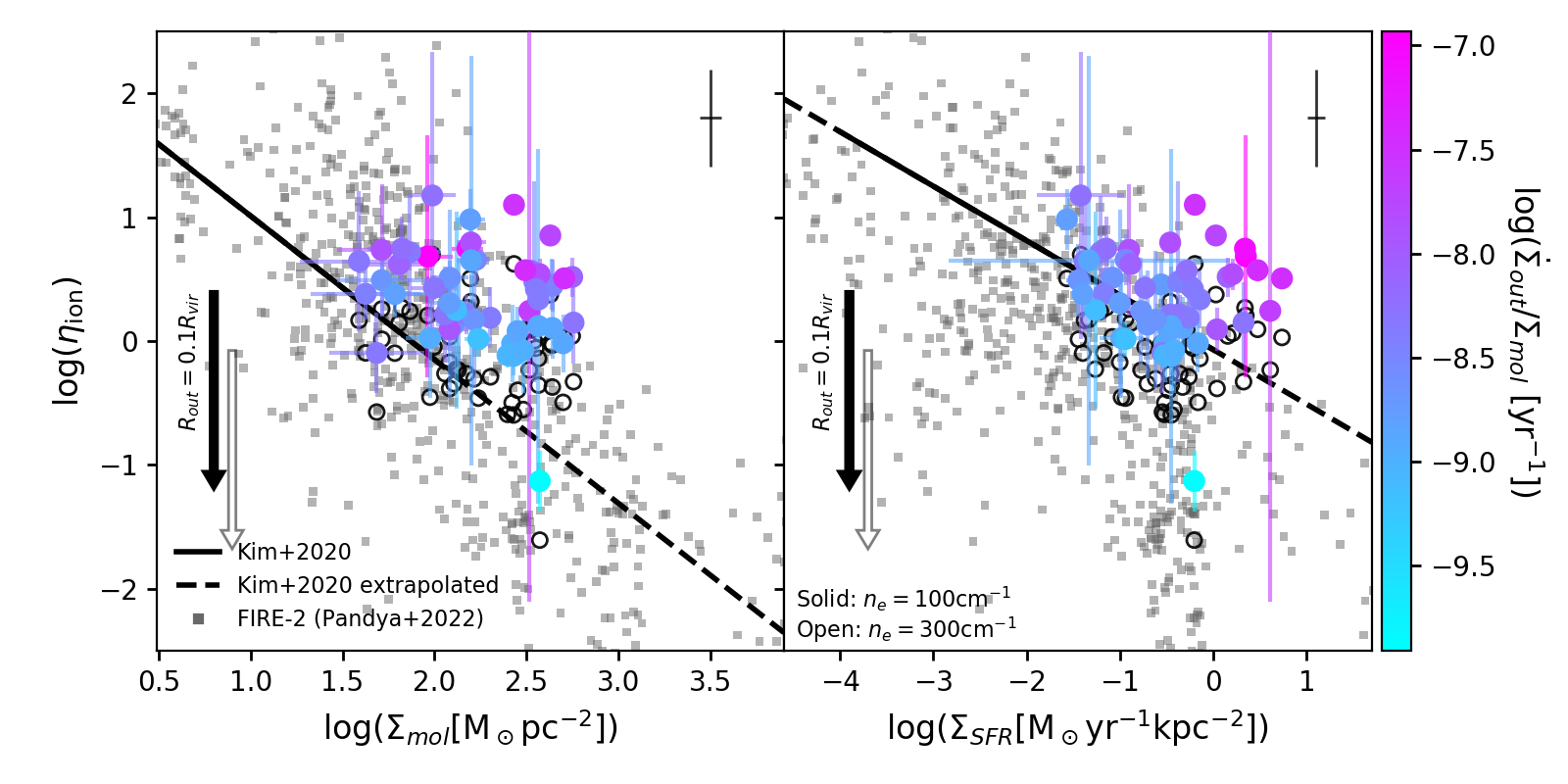}
    \caption{The mass loading factor of ionised gas, $\eta_\mathrm{ion}$, as a function of the molecular gas surface density, \sigmamol\ (left panel) and the star formation rate surface density, \sigmasfr\ (right panel).  Observational data from IRAS08 is given in circles coloured by outflow efficiency using $n_e=100$~cm$^{-1}$, with rescaled values for an electron density of $n_e=300$~cm$^{-1}$ as open circles.  The black arrows show the magnitude and direction the median $\eta_{\rm ion}$ value would move if we assumed an outflow extent of $R_{\rm out}=0.1R_{\rm vir}$ (open arrow represents the median with $n_e=300$~cm$^{-1}$).  Typical error bars for the IRAS08 data are shown in the top right corner of each panel. We compare our observational results from IRAS08 with results from the FIRE-2 \citep[][Figure 11]{pandya2021characterisingFIRE2} and TIGRESS \citep[][Figure 9]{kim2020firstresultssmaug} simulations.  Results from the warm-phase gas ($10^3<T<10^5$~K) from FIRE-2 are shown as grey squares. The \sigmamol-$\eta_\mathrm{ion}$ and \sigmasfr-$\eta_\mathrm{ion}$ relationships found in \cite{kim2020firstresultssmaug} for the cool gas component ($T<2\times10^4~$K) are plotted with solid black lines, and extrapolated past the parameter space they covered with the dashed black lines.  The simulation data is not inconsistent with our observational results.}
    \label{fig:FIRE2_comparison}
\end{figure*}

In Figure~\ref{fig:FIRE2_comparison} we compare the mass loading factor, $\eta_{\rm ion}=\dot{\Sigma}_{\rm out}/\Sigma_{\rm SFR}$, to  $\Sigma_{\rm mol}$ and $\Sigma_{\rm SFR}$ in IRAS08.  We additionally overplot results from the recent FIRE-2 simulations \citep{pandya2021characterisingFIRE2} and the SMAUG-TIGRESS simulation \citep{kim2020firstresultssmaug}. A number of simulations and analytic theories predict a decreasing $\eta_{\rm ion}$ with increasing $\Sigma_{\rm SFR}$ \citep[recently][]{fielding2017SNelaunch, li2017supernovaedriven,kim2020firstresultssmaug,pandya2021characterisingFIRE2}. Observations are mixed as to how well these correlate  \citep{arribas2014ionisedgasoutflows,robertsborsani2020outflowsMANGA}.

In IRAS08 we find very little correlation between $\eta_{\rm ion}$ and $\Sigma_{\rm mol}$.  We find a Pearson correlation coefficient of $r=-0.23$ (p-value $=0.1$) for $\log(\Sigma_{\rm mol})-\log(\eta_{\rm ion})$. We note, however, that values of $\Sigma_{\rm mol}$ in IRAS08 cover a fairly small range compared to the FIRE-2 simulations. We also note that the FIRE-2 simulations calculate the total values of $\eta$ and $\Sigma_{\rm mol}$ for entire galaxy halos, rather than resolved regions.  Moreover, the values of IRAS08, while being on the high side of the FIRE-2 data and over the \citet{kim2020firstresultssmaug} correlation, are not completely discrepant. Nevertheless, our data are consistent with no correlation between $\eta$ and $\Sigma_{\rm mol}$. This is in contrast to the steeper trend expected from both the TIGRESS and FIRE-2 simulations \citep{kim2020firstresultssmaug, pandya2021characterisingFIRE2}.  It could be that we should expect different relations when comparing our resolved observations to simulations on different spatial scales.  In order to recover the high scatter correlation which simulations predict between $\eta_{\rm ion}$ and $\Sigma_{\rm mol}$, we may need to include observations of more galaxies covering a larger range in $\Sigma_{\rm mol}$.

For $\log(\Sigma_{\rm SFR})-\log(\eta_{\rm ion})$ we similarly find a weak-to-no trend (Pearson correlation coefficient of $r=-0.17$, p-value $=0.23$). We note that below $\log(\Sigma_{\rm SFR} [$M$_{\odot}$~yr$^{-1}$~kpc$^{-2}])\approx -0.5$ the data in IRAS08 is much more consistent with theoretical expectations from \cite{kim2020firstresultssmaug} and is consistent with a negative correlation (correlation coefficient of $r=-0.67$, p-value $=2.5\times10^{-4}$). It is possible that regions of the galaxy below this $\Sigma_{\rm SFR}$ follow the predicted relationship, but that in regions with higher $\Sigma_{\rm SFR}$ the outflow efficiency increases (see Fig.~\ref{fig:outeff}) and we therefore observe a flat $\eta_{\rm ion}$.  It is important to also point out that the simulations of \cite{kim2020firstresultssmaug} are only run to $\Sigma_{\rm SFR}\sim1.0$~M$_{\odot}$~yr$^{-1}$~kpc$^{-2}$ and to  $\Sigma_{\rm mol}\sim100$~M$_{\odot}$~pc$^{-2}$. We have extrapolated their \sigmamol-$\eta_{\rm ion}$ and \sigmasfr-$\eta_{\rm ion}$ relations past this parameter space in order to compare to our observations of a starbursting environment.  It is unclear whether this extrapolation should hold in starbursting environments.

We note that the systematic uncertainty of $n_e$ is such that assuming higher values of $n_e$ would bring our observed $\Sigma_{\rm mol}-\eta_{\rm ion}$ into better agreement with the \cite{kim2020firstresultssmaug} prediction. This would, however, simultaneously make the mass-loading factors too low in comparison to $\Sigma_{\rm SFR}$. 

While we observe the molecular gas surface density \sigmamol, the TIGRESS simulations measure the total ISM gas mass divided by the horizontal area of the disk included in the box of the simulation.  Their horizontal box sizes range from 512~pc to 2048~pc.  The largest of these is $\sim4\times$ the area of our pixels.
Moreover, we need to be mindful that the comparison to the FIRE-2 results is not an apples-to-apples comparison. \cite{pandya2021characterisingFIRE2} defined particles as outflows if they flow away from the galaxy disk and reach $0.1R_{\rm vir}$ with enough energy remaining to make it to a larger radius.  To convert the down-the-barrel observations of outflows into a mass outflow rate we assume an ``outflow radius" of $R_{\rm out}=500~\mathrm{pc}$ for our target \citepalias[for more details, see][]{reichardtchu2022resolvedmaps}.  This is four times smaller than the $0.1R_{\rm vir}$ used by \cite{pandya2021characterisingFIRE2}, which would be $\sim20~\mathrm{kpc}$ for a galaxy of IRAS08's mass.
We don't know how the velocity and density profiles of the outflow change with radius, however assuming everything stays the same, we rescale our data to use $R_{\rm out}=20~\mathrm{kpc}$. Rescaling our data decreases the median value of $\eta_{\rm ion}$ from 2.5 to 0.06. This rescaling introduces significant uncertainty. It would be useful in the future for simulations to estimate the properties of outflows such as the mass loading factor of the warm ionised gas in the region less than 10~kpc from the galaxy disk, where we observationally measure outflow properties.

\subsection{Breakout Outflows} 
\label{subsec:breakout_outflows}

\begin{figure}
    \centering
    \includegraphics[width=1.\linewidth]{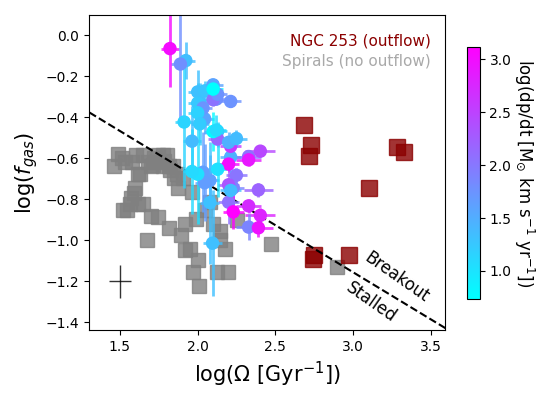}
    \caption{The gas fraction, $f_{\rm gas}\equiv \Sigma_{\rm mol}/(\Sigma_{\rm mol}+\Sigma_{\rm star})$, is compared to the orbital frequency, $\Omega$.  Data points for IRAS08 are coloured by the outflow momentum flux, $dp/dt$.  A typical error bar for the IRAS08 data is shown in the bottom left corner. The dashed line shows the boundary predicted by \cite{orr2022burstingbubbles_letter} between outflows which break out of the disk, and outflows which are stalled within the disk (see their Figures~1 and 2).  Grey squares indicate resolved observations of local spiral galaxies with no observed outflow.  Red squares indicate the resolved observations of the outflow in NGC~253.  We find that $\sim85\%$ of our resolved regions are within the predicted region for breakout outflows.}
    \label{fig:fgasomega}
\end{figure}

In a series of papers, \cite{orr2021burstingbubbles_long,orr2022burstingbubbles_letter} considered the impacts of supernova clustering on the regulation of star formation by stellar feedback, and the conditions in which supernovae leave the disk or impart their momentum to the ISM. They outlined the scenarios in which a resulting superbubble generated by a supernova explosion expands until it either breaks out of the galaxy disk, or stalls within the galaxy. This simple, yet predictive, model is motivated by and is consistent with simulations of outflows \citep{fielding2018clustered} and star cluster formation \citep{grudic2018whenfeedbackfails}.  As the authors define it, in the event of ``breakout", more than 60\% of the momentum from the subsequent supernovae transfers to the outflow rather than coupling to the ISM.

\cite{orr2021burstingbubbles_long} put forward the prediction that a boundary line exists in the parameter space between local gas fraction, $f_{\rm gas}\equiv\Sigma_{\rm gas}/(\Sigma_{\rm gas}+\Sigma_{\rm star})$, and the orbital frequency, $\Omega\equiv v(R)/R$, that determines whether an outflow travels beyond the vertical scale-height of the disk (i.e.~breakout) or stays bound inside the disk (i.e.~stalled)  \citep[see also][]{orr2022burstingbubbles_letter}. 

We compare our observations for IRAS08 to the predictions from \cite{orr2022burstingbubbles_letter} in Figure~\ref{fig:fgasomega}.  We find that the vast majority  ($\sim85\%$) of our resolved regions in IRAS08 have $f_{\rm gas}$ and $\Omega$ consistent with the predicted region for breakout outflows. We take the velocity models used in \citetalias{fisher2022extremevariation} for IRAS08, in which a flat rotation curve is fit to the CO(2-1) data, such that $v(R)=v_{\rm flat}[1-\exp(-R/r_{\rm flat})]$. There is a significant systematic uncertainty in IRAS08 due to the low inclination of the galaxy. We adopt $i\approx20^\circ$, which \citetalias{fisher2022extremevariation} derived from the HST/F550M isophotes at large radius, which is similar to the assumed inclination in \cite{leitherer2013FUVobservations}. An uncertainty in the inclination angle of $\pm 5^\circ$ would result in an uncertainty in Figure~\ref{fig:fgasomega} of $\pm0.15$~dex in $\log(\Omega)$. Observationally this produces a challenging balance, as outflows would become more difficult to observe in more highly inclined systems. The figure also compares the IRAS08 measurements to NGC~253, which is known to contain an outflow. Given that we have thus far seen similar behaviour between NGC~253 and IRAS08, it is therefore further consistent that both are above the breakout line. Moreover, as a control \citet[][see their Fig.~1]{orr2022burstingbubbles_letter} showed data for 4~spiral galaxies that are known to not have outflows, which we show are not colocated with IRAS08. 

We can compare the outflow velocity in IRAS08 to local velocity dispersion to check if the \cite{orr2021burstingbubbles_long, orr2022burstingbubbles_letter} breakout scenario is consistent with outflows moving beyond the scale-height of the disk. The scale-height of a disk is set by $h_z \propto \sigma^2/\Sigma_{\rm gas}$, where $\sigma$ is the velocity dispersion \citep[see discussion in][]{wilson2019kennicuttschmidtlaw}. For standard scenarios, if an outflow is moving faster than $\sigma$ it will reach the scale-height of the disk. \citetalias{fisher2022extremevariation} showed that the velocity dispersion of the molecular gas is on average 25$\pm$6~km~s$^{-1}$, with the highest values being $\sim$40~km~s$^{-1}$. The minimum of our estimated outflow velocities, $v_{\rm out}$, is of order 130~km~s$^{-1}$, which is well above the dispersion of the disk, even if we account for systematic differences between ionised and molecular gas velocity dispersions \citep{girard2021systematic}. The position of IRAS08 in the $f_{\rm gas}-\Omega$ diagram is therefore consistent with the theory put forward in \cite{orr2021burstingbubbles_long}. We can also argue that this implies the depletion of gas discussed in previous sections is indeed leaving the plane of the disk, and removing gas from star formation.

\cite{orr2021burstingbubbles_long} also predicted that at fixed $f_{gas}$ and larger $\Omega$ less momentum couples to the gas in the disk, and more momentum couples to the outflow.  
For every unit of new stellar mass which is formed, there is an amount of momentum which is available to be coupled to the surrounding gas (ISM or outflow).  
We do not know the exact relationship between our observable outflow momentum flux, $dp/dt$, and the amount of momentum per unit new stellar mass formed, $p/m_*$. 
While we cannot observe the momentum which couples to the ISM gas, we can observe the momentum flux of the outflow in each region.  
In Figure~\ref{fig:fgasomega} we colour the  points of IRAS08 by their outflow momentum flux.  
We find that increasing $\Omega$ corresponds with increasing momentum flux within the outflow, i.e. outflows launched closer to the galaxy center have greater momentum. Note that to first order $dp/dt$ simply traces $\dot{M}_{\rm out}$, we therefore expect similar correlations with $\dot{\Sigma}_{\rm out}$. 
There are a number of uncertainties, not the least of which is likely the phase distribution of the momentum of the outflow (see \citealp{fielding2022structure} for an in-depth theoretical exploration of the momentum content of multiphase galactic winds). Nonetheless, it is plausible that as more momentum flux is incorporated in the outflow this would imply that less is available to generate turbulence in the ISM. The effect is thus to decrease the effective $p/m_*$ felt by the disk surrounding the outflow.

\section{Summary \& Discussion}
\label{sec:summary}

\subsection{Summary}
We present resolved measurements of IRAS08 in ionised gas using KCWI/Keck and molecular gas using NOEMA.  We use these observations to relate the molecular gas mass and star formation observed in the galaxy to the resolved outflow and its properties. We have shown that the combined comparison of $\Sigma_{\rm SFR}$, $\Sigma_{\rm mol}$ and $\dot{\Sigma}_{\rm out}$ allows for very direct comparison to theory, in which three parameters that are thought to combine to regulate star formation in disk galaxies are characterised.        

We have shown direct correlation between the resolved $\Sigma_{\rm SFR}$ and the co-located outflow mass flux, $\dot{\Sigma}_{\rm out}$, in regions of $\sim$500~pc. We find that  $\dot{\Sigma}_{\rm out}$ correlates much more strongly with $\Sigma_{\rm SFR}$ than $\Sigma_{\rm mol}$. This leads to a connection between outflow strength and location in the Kennicutt-Schmidt relationship between $\Sigma_{\rm SFR}-\Sigma_{\rm mol}$, such that regions of the galaxy with shorter molecular gas depletion times have stronger outflows. There is therefore a strong, superlinear relationship between the outflow efficiency $\dot{\Sigma}_{\rm out}/\Sigma_{\rm mol}$ and the inverse of the gas depletion time $t_{\rm dep}^{-1}=\Sigma_{\rm SFR}/\Sigma_{\rm mol}$.

We find that the mass-loading factors we observe are consistent with predictions from multiple simulations \citep{pandya2021characterisingFIRE2,kim2020firstresultssmaug}. 
The outflows we resolve are consistent with breakout outflows according to the region in the $f_{\rm gas}-\Omega$ plane defined by \cite{orr2021burstingbubbles_long, orr2022burstingbubbles_letter}. For our galaxy we can compare the velocity of the outflow to the local velocity dispersion, which we indeed find is consistent with outflows travelling fast enough to leave the disk.

\subsection{Systematic Uncertainties}
In this paper we attempt to derive physical quantities of outflows from observations. This is necessary to compare to theory and simulation, however the derivation is heavily impacted by assumptions that introduce systematic uncertainties. We do not have strong constraints on how the $n_e$ of the outflow changes across either a range of galaxies, or for regions within a galaxy. Moreover, outflows are clearly multiphase phenomena \cite[e.g][]{fluetsch2021propertiesmultiphase,fluetsch2019coldmolecularoutflows, herrera-camus2020AGNfeedback}. However, the ratio of the mass outflow rate in ions to other phases of gas is not well constrained by observations, especially on resolved scales within galaxies. \cite{leroy2015multiphaseM82} provided a heuristic model in which this ratio could change based on the local region of the galaxy. More work constraining these properties with ALMA, MUSE and KCWI is direly needed in order to reduce these systematics, and confirm the results we have presented.

\subsection{Implications for star formation regulation in high \texorpdfstring{$\Sigma_{\rm SFR}$}{TEXT} galaxies}
The superlinear relationship between outflow efficiency and star formation efficiency (Fig.~\ref{fig:outeff}) has implications for how galaxies regulate their star formation. This correlation suggests that in the disk-mode of star formation, the outflow and the star formation are similarly effective at removing gas, thus regulating the ability to form new stars. However, as the disk moves into a starburst mode, this regulation becomes more dominated by the outflow. 

We can therefore outline a picture in which the nature of the mechanism regulating star formation may change with respect to location in the $\Sigma_{\rm SFR}-\Sigma_{\rm mol}$ plane. At a fixed $\Sigma_{\rm mol}$, as the $\Sigma_{\rm SFR}$ increases above the typical depletion time of $\sim1-2$~Gyr, the outflows become the dominant mechanism in removing gas. \cite{krumholz2018unifiedmodel} argues that gas inflow rates in disks are comparable to the SFR, and therefore if the mass-outflow rate becomes significantly greater than $\eta\sim1$, as we show in Fig.~\ref{fig:FIRE2_comparison}, then the gas removal by the outflow will reduce the gas surface density. This then causes the star formation rate surface density to decline, and further gas removal is dominated by star formation rather than outflows. Outflows in the center of starbursting disks, such as observed here, may therefore act to reduce the impacts of other effects that drive up the gas velocity dispersion. Theoretical work that incorporates outflows into dynamical equilibrium models of star formation may be necessary to explain high $\Sigma_{\rm SFR}$ disk galaxies. 

We note that this is heuristically similar to a scenario in which feedback can be overcome on short timescales by very efficient star formation \citep{torrey2017instability}. \citet{orr2019simplenon-equilibrium} and \citet{rathjen2021SILCCVImultiphaseISM} made similar analyses of simulation data, at lower $\Sigma_{\rm mol}$. The difference between IRAS08 and the conclusions in \cite{torrey2017instability}, however, is that in \cite{torrey2017instability} this phenomenon only occurs over the very central nuclei of a galaxy, whereas in IRAS08 it extends to beyond the half-light radius of the star-light. \citetalias{fisher2022extremevariation} reported short free-fall times, $\sim1-5$~Myr in a very large fraction of the galaxy, and high star formation efficiencies per free-fall time. Taken together, the observations of \citetalias{fisher2022extremevariation} and the outflow observations here are likewise conceptually consistent with the picture in which departures from dynamical equilibrium are regulated by strong winds.

We note that there are many similarities between IRAS08 and $z\sim1-2$ galaxies, including the higher gas fraction, high molecular gas velocity dispersion and location in the $\Sigma_{\rm SFR}-\Sigma_{\rm mol}$ diagram \citepalias[for a full comparison see][]{fisher2022extremevariation}. Our results may therefore indicate that the low molecular gas depletion times observed in galaxies at $z\sim2$ \citep[e.g.][]{tacconi2018PHIBSSunified, herrera-camus2019molecularionisedgas} may be due more to outflows depleting the local area of the disk, rather than conversion of gas into stars. What we do not currently know is which relationship is more important, that of $\dot{\Sigma}_{\rm out}-\Sigma_{\rm SFR}$ or the relationship between outflow efficiency and star formation efficiency. For example, both \cite{genzel2011sins} and \cite{molina2019kpcscalegaskinematics} found galaxies with high $\Sigma_{\rm mol}$ and $\Sigma_{\rm SFR}$ at sub-galactic resolution, but they have disk-like depletion times. Observations of outflows in systems like this would be informative.

Our work has shown the diagnostic power of comparing the properties of outflows to both $\Sigma_{\rm SFR}$ and $\Sigma_{\rm mol}$ in resolved observations. 
Tracking three of the parameters responsible for regulating star formation in galaxy disks in resolved observations enables us to make unique comparisons to theory.
Yet we have examined only one galaxy.
To repeat this on multiple targets, however, requires a substantial investment of observing time. Outflows are more easily detected on high efficiency spectrographs on 8-10~m class telescopes (such as Keck/KCWI and VLT/MUSE). This then must be combined with data from millimeter-wave interferometers, such as ALMA, SMA and NOEMA. 
Finally, observations of the ratio of $\dot{M}_{\rm out}({\rm ions})$-to-$\dot{M}_{\rm out}({\rm molecules})$ vary significantly on the handful of targets that have been measured \citep{fluetsch2019coldmolecularoutflows, herrera-camus2020AGNfeedback}. Combining resolved observations from optical and millimeter-wave instruments, we would ideally have much better constraints on the phase distribution of outflows. Deep ALMA observations of molecular gas outflows targeting galaxies with known ionised gas outflows would be very informative, as ionised gas outflows are much easier to observe, especially in more distant galaxies. We expect future work using DUVET galaxies to address these concerns.


\vspace{0.5cm}
We are grateful to Viraj Pandya, for discussion and sharing FIRE-2 simulation results, and to Matt Orr for helpful discussion. Parts of this research were supported by the Australian Research Council Centre of Excellence for All Sky Astrophysics in 3 Dimensions (ASTRO 3D), through project number CE170100013.  
D.B.F. acknowledges support from Australian Research Council (ARC) Future Fellowship FT170100376 and ARC Discovery Program grant DP130101460. 
A.D.B. acknowledges support from NSF-AST2108140.
A.J.C. acknowledges funding from the ``FirstGalaxies'' Advanced Grant from the European Research Council (ERC) under the European Union's Horizon 2020 research and innovation programme (Grant agreement No. 789056).
R.H.-C. thanks the Max Planck Society for support under the Partner Group project ''The Baryon Cycle in Galaxies" between the Max Planck for Extraterrestrial Physics and the Universidad de Concepción. R.H-C also acknowledges financial support from Millenium Nucleus NCN19058 (TITANs) and support by the ANID BASAL projects ACE210002 and FB210003.
D.O. is a recipient of an Australian Research Council Future Fellowship (FT190100083) funded by the Australian Government. 
R.R.V. and K.S. acknowledge funding support from National Science Foundation Award No. 1816462.

This work is based on observations carried out under project number W17CB with the IRAM NOEMA Interferometer. IRAM is supported by INSU/CNRS (France), MPG (Germany) and IGN (Spain).

Some of the data presented herein were obtained at the W. M. Keck Observatory, which is operated as a scientific partnership among the California Institute of Technology, the University of California and the National Aeronautics and Space Administration. The Observatory was made possible by the generous financial support of the W. M. Keck Foundation. Observations were supported by Swinburne Keck program 2018A\_W185. The authors wish to recognise and acknowledge the very significant cultural role and reverence that the summit of Maunakea has always had within the indigenous Hawaiian community. We are most fortunate to have the opportunity to conduct observations from this mountain.

%

\vspace{5mm}
\facilities{Keck:II (KCWI), IRAM:NOEMA}


\software{koffee \citep[][\url{https://github.com/bronreichardtchu/koffee/tree/PaperI-code}]{reichardtchu2022resolvedmaps},  
        astropy \citep{astropy2013, astropy2018}
          }





\bibliography{bibliography}{}
\bibliographystyle{aasjournal}



\end{document}